\documentclass[nofootinbib,notitlepage]{article}

\usepackage{amsmath}
\usepackage{amssymb}
\usepackage[pdftex]{graphicx}
\usepackage[english]{babel}
\usepackage{authblk}

\begin{document}
\title{Constraints on Covariant Horava-Lifshitz Gravity from frame-dragging experiment}

\author{Ninfa Radicella\thanks{ninfa.radicella@sa.infn.it}}
%\email{ninfa.radicella@sa.infn.it}
\author{Gaetano Lambiase\thanks{lambiase@sa.infn.it}}
%\email{lambiase@sa.infn.it}
\author{Luca Parisi\thanks{parisi@sa.infn.it}}
%\email{parisi@sa.infn.it}
\author{Gaetano~Vilasi\thanks{vilasi@sa.infn.it}}
%\email{vilasi@sa.infn.it}

\affil{Dipartimento di Fisica ``E.R. Caianiello", Universit\`{a} di Salerno,\\ Via Giovanni Paolo II  132, 84081 Fisciano (Sa), Italy\\
and \\
 INFN, Sezione di Napoli, Gruppo Collegato di Salerno, 80126~Napoli, Italy}

\maketitle

\begin{abstract}
The effects of Horava-Lifshitz corrections to the gravito-magnetic field are analyzed.
Solutions in the weak field, slow motion limit, referring to the motion of a satellite around the Earth are considered. The post-newtonian paradigm is used to evaluate constraints on the Horava-Lifshitz parameter space from current satellite and terrestrial experiments data.
In particular, we focus on GRAVITY PROBE B, LAGEOS and the more recent LARES mission,
as well as a forthcoming terrestrial project, GINGER.
\end{abstract}

%%%%%%%%%%%%%%%%%%%%%%%%%%%%%%%%%%%%%%%%%%%%%%%%%%%%%%%%%%%%%%%%%%%%%%%%%%%%%%%%%%%%%%%%%%%%%%%%%%%%%%%%%%%%%%%%%%%%%%%%%%%%%%%%%%%%%%%%%%%%%%%%%%%%%%%%%%%%%%%%%%%%%%%%%%%%%%%%%%
\section{Introduction}
 Tests on gravitational theories alternative to General Relativity, at a fundamental level,  are essential to explore possible generalisations of Einstein theory \cite{will}. General Relativity has been deeply tested at Solar System level hence any alternative theory of gravity should pass such tests, basically reproducing General Relativity at the Solar System scale \cite{ruggiero07, smith08, bohmer10,harko11,lambiase13}.
 
 Among these theories, Horava-Lifshitz gravity is gaining much attention in recent years. Horava-Lifshitz is a power counting renormalisable theory based on an anisotropic scaling of space and time in the ultraviolet limit \cite{Horava09}. Diffeomorphism invariance is thus replaced by the so-called foliation-preserving diffeomorphism, and the theory violates the local Lorentz invariance in the ultraviolet. Such a symmetry is expected to be recovered in the infrared (IR) limit thanks to the renormalization group flow for the couplings of the model but still no results support this specific behaviour. Nevertheless, this alternative model of gravity with its healthy ultraviolet behaviour has attracted a big interest and has been analysed both in cosmology and in the weak field limit. Unfortunately, the breaking of general covariance introduces a dynamical scalar mode that may lead to strong coupling problem and instabilities \cite{wang11}.\\
 Recently, a new Covariant version of Horava Lifshitz gravity has been formulated by Horava and Melby-Thompson \cite{HL} which includes two additional nondynamical fields $A$ and $\phi$, together with a new $U(1)$ symmetry. The latter eliminates the new scalar degree of freedom thus curing the strong coupling problem in the IR limit.\\
In this paper we refer to the covariant version of Horava and Melby-Thompson with the coupling $\lambda$, in the extrinsic curvature term of the action, not forced to be $1$ as presented in \cite{daSilva11}.\\
In order to constrain such a theory against Solar System tests we will use the Parametrized Post Newtonian framework (PPN). In this formalism the metric of an alternative metric theory of gravity is analyzed in the weak field and slow motion limit and its deviations from General Relativity are expressed in terms of PPN parameters \cite{willbook}. Once a metric has been obtained, one can calculate predictions of the alternative theory which actually depend on these PPN parameters. The above-mentioned approximation allows to describe the spacetime around a spinning mass. Here, we look at the effects on the orbits of test particles and the precession of  spinning objects in this spacetime. 
In particular we focus on the results on the de Sitter (geodetic precession) \cite{desitter16} and the Lense-Thirring (frame dragging) \cite{lense18} effects. Experimental measurements of such physical effects directly lead to constraints on the parameters  of Horava-Lifshitz theory \cite{wheeler}.   We will consider the results of two completed space experiments: Gravity Probe B (GP-B) \cite{gpb} and Lageos \cite{lageos}. Moreover, we will comment on the expected results of LARES \cite{lares} and present what will come from GINGER \cite{ginger}, an Earth based experiment.\\
The paper is organized as follows. In section \ref{HLmodel} we present the covariant version of the Horava-Lifshitz gravity. Then, in sec. \ref{Weak field approximation and spherically symmetric solution} we focus on the weak field and slow motion approximation and present the solution in the PPN formalism, as in \cite{lin13}. In particular we derive the predictions of the theory on the geodetic precession and the frame dragging effect that are compared and contrasted with the experimental results from space experiments in sec. \ref{Constraints  from space experiments}. In the subsequent section we present the expected results by a terrestrial experiment, GINGER, and finally sec. \ref{Conclusions}  contains our conclusions. 

%%%%%%%%%%%%%%%%%%%%%%%%%%%%%%%%%%%%%%%%%%%%%%%%%%%%%%%%%%%%%%%%%%%%%%%%%%%%%%%%%%%%%%%%%%%%%%%%%%%%%%%%%%%%%%%%%%%%%%%%%%%%%%%%%%%%%%%%%%%%%%%%%%%%%%%%%%%%%%%%%%%%%%%%%%%%%%%%%%%%%%%%%%%%%%%%%%%%%%%%%%%%%%%
\section{HL model}\label{HLmodel}

One of the candidates for quantum gravity is Horava-Lifshitz theory which is power-counting renormalizable due to the anisotropic scaling of space and time \cite{Horava09} in the ultra-violet limit,
$$
t\rightarrow\sl{l}^3 t, \quad \quad \vec{x}\rightarrow\sl{l}  \vec{x}.
$$

According to the Hamiltonian formulation of General Relativity developed by Dirac  \cite{dirac} and Arnowitt Deser and Misner \cite{arnowitt59}, the suitable variables in this Horava-Lifshitz theory are the lapse function, the shift vector and the spatial metric $N$, $N_i$, $g_{ij}$ respectively, so that
$$
ds^2=-N^2 dt^2+g_{ij} \left(dx^i+N^i dt\right)\left(dx^j+N^j dt\right).
$$
The gauge symmetry of the system is the foliation-preserving diffeomorphism $Diff(M,\mathcal{F})$,
\begin{equation}\label{diff}
\tilde{t}=t-f(t),\quad \tilde{x}^i=x^i-\zeta^i(t,x),
\end{equation}
which causes a non-healthy new degree of freedom in the gravitational sector, a spin-0 graviton. This scalar mode is not stable on a Minkowski background neither in the original version of the Horava-Lifshitz theory \cite{Horava09} nor in the Sotiriou, Visser and Weinfurtner implementation \cite{SVW09}. The problem is ameliorated when de Sitter spacetime is considered, but the strong coupling problem  still exists. Infact, being the extra degree of freedom always coupled, General Relativity is not recovered in the perturbative limit. \\
A generalisation of the original version of the Horava-Lifshitz theory has been recently proposed  \cite{HL,daSilva11}, in which the spin-0 mode has been eliminated from the theory by extending the foliation-preserving-diffeomorphism to include a local $U(1)$ symmetry. This approach allows to restore the general covariance.\\
 In order to heal the scalar graviton problem, two new fields have been introduced: the gauge field $A(t,x)$ and the Newtonian pre-potential $\phi(t,x)$.
The theory satisfies the projectability condition, i.e. the lapse function only depends on time $N=N(t)$, while the total gravitational action is given by
\begin{equation}\label{HLaction}
S_g=\zeta^2\int dt\  d^3x\  N\sqrt{g}\left(\mathcal{L}_K-\mathcal{L}_V+\mathcal{L}_\phi+\mathcal{L}_A\right),
\end{equation}
where $g=\text{det}(g_{ij})$ and 
\begin{eqnarray*}
\mathcal{L}_K&=&K_{ij} K^{ij} -\lambda K^2,\\
\mathcal{L}_\phi&=&\phi\ \mathcal{G}^{ij}\left(2K_{ij}+\nabla_i\nabla_j \phi\right),\\
\mathcal{L}_A&=&\frac{A}{N}\left(2 \Lambda_g-R\right).
\end{eqnarray*}
Covariant derivatives as well as Ricci terms all refer to the $3$-metric $g_{ij}$.\\
$K_{ij}$ represents the extrinsic curvature
$$
K_{ij}=g_i^k\nabla_k n_j,
$$
$n_j$ being a unit normal vector of the spatial hypersurface,  and $\mathcal{G}_{ij}$  the $3$-~dimensional generalised Einstein tensor
$$
\mathcal{G}_{ij}=R_{ij}-\frac{1}{2}g_{ij} R+\Lambda_g g_{ij}.
$$
 We remark that  $\lambda$ characterises deviations of the kinetic part of the action from General Relativity.
The most general parity-invariant Lagrangian density up to six order in spatial derivatives is
\begin{eqnarray*}
\mathcal{L}_{V}&=&2 \Lambda-R+\frac{1}{\zeta^2}\left(g_2\  R^2+g_3\  R_{ij} R^{ij}\right)\\
&&+\frac{1}{\zeta^4}\left(g_4\  R^3+g_5\  R R_{ij} R^{ij}+g_6\  R^i_j R^j_k R_i^{k}\right.\\
&&\left.+g_7\  R \nabla^2 R+ g_8\  (\nabla_i R_{jk}) (\nabla^i R^{jk})\right),
\end{eqnarray*}
where in physical units $\zeta^2=(16 \pi G)^{-1}$,  $G$ being the Newtonian constant in the Horava-Lifshitz theory.\\
Matter coupling for this theory has not been studied systematically; in \cite{lin12} it has been shown that, in order to be consistent with solar system tests, the gauge field and the Newtonian pre-potential must be coupled to matter in a specific way, but there were no indication on how to obtain the precise prescription from the action principle. Recently such a prescription has been generalised \cite{lin13} and a scalar-tensor extension of the theory presented above has been developed to allow the needed coupling to emerge in the IR without spoiling the power-counting renormalizability of the theory in the ultraviolet. In details,  the matter action term is    
\begin{equation}
S_M=\int dt d^3 x\tilde{N}\sqrt{\tilde{g}}\  \mathcal{L}_M\  (\tilde{N}, \tilde{N}_i,\tilde{g}_{ij}; \psi_n),
\end{equation}
where $\mathcal{L}_M$ is the matter Lagrangian and $\psi_n$ stands for matter fields.The metric is given by
\begin{equation}
(\gamma_{\mu\nu})=\left(
\begin{array}{cc}
-\tilde{N}^2+\tilde{N}^i\tilde{N}_i &\tilde{N}_i \\ \tilde{N}_i & \tilde{g}_{ij}.
\end{array}
\right)
\end{equation}
This means that matter fields couple to the Arnowitt-Deser-Misner components with the tilde, defined as
\begin{eqnarray*}
\tilde{N}&=&(1-a_1\sigma) N,\\
 \tilde{N}^i&=&N^i+N g^{ij}\nabla_j\phi,\\
 \tilde{g}_{ij}&=&(1-a_2\sigma)^2 g_{ij},
\end{eqnarray*}
where $a_1$ and $a_2$ are two arbitrary coupling constants and
 $$
 \sigma=\frac{A-\mathcal{A}}{N}, \quad\text{with}\quad\mathcal{A}=-\dot{\phi}+N^i\nabla_i\phi+\frac{1}{2}N\nabla^i\phi \nabla_i\phi.
 $$
%%%%%%%%%%%%%%%%%%%%%%%%%%%%%%%%%%%%%%%%%%%%%%%%%%%%%%%%%%%%%%%%%%%%%%%%%%%%%%%%%%%%%%%%%%%%%%%%%%%%%%%%%%%%%%%%%%%%%%%%%%%%%%%%%%%%%%%%%%%%%%%%%%%%%%%%%%%%%%%%%%%%%%%%%%%%%%%%%%%%%%%%%%%%%%%%%%%%%%%%%%%%%%%
\section{Weak field approximation and spherically symmetric solution}\label{Weak field approximation and spherically symmetric solution}
In order to find a viable solution for solar system constraints, we assume that the influence of the cosmological constant and the space curvature are negligible; hence
\begin{equation}
\Lambda=\Lambda_g=0.
\end{equation}
Furthermore, in the IR all the higher order derivative terms are small, then we can safely set $g_2,\dots, g_8$ to zero in what follows. A caveat must be clarified. The coupling $g_1$ cannot be rescaled to $g_1=-1$, that corresponds to the General Relativity value and represents a redefinition of the units of time and space. Here, this freedom has already been used to set to unity other parameters that enter in the matter coupling \cite{lin13}. In addition, we can use the $U(1)$ gauge freedom to choose $\phi=0$, which uniquely fixes the gauge.\\
We consider the metric in the post-Newtonian approximation; it can be written in the form \cite{will}
\begin{equation}
\gamma_{\mu\nu}=\eta_{\mu\nu}+h_{\mu\nu},
\end{equation}
where $\eta_{\mu\nu}=\text{diag} (-1,1,1,1)$ and we consider the perturbation up to third order since we are interested in testing post-newtonian effect on a satellite orbiting around the Earth:
\begin{eqnarray}
h_{00}&\sim& \mathcal{O}(2)\nonumber\\
h_{0i}&\sim&\mathcal{O}(3)\nonumber\\
h_{ij}&\sim&\mathcal{O}(2),
\end{eqnarray}
where $\mathcal{O}(n)\equiv\mathcal{O}(v^n)$, $v$ being the three velocity of the objects considered (we are using natural units).\\
Following \cite{lin13}, the solution in the Horava-Lifshitz theory, up to the above-mentioned order, is expressed by
\begin{eqnarray}\label{PPN}
h_{00}&\sim& 2U\nonumber\\
h_{0i}&\sim&c V_i+d \chi_{,0i}\nonumber\\
h_{ij}&\sim&2\gamma U\delta_{ij},
\end{eqnarray}
where the gauge freedom has been used to eliminate anisotropic terms in the space-space contribution of the perturbation. The coefficients are found by solving the equations at the appropriate order and can be expressed in terms of the couplings that appear in the matter and gravitational Horava-Lifshitz action. They read
\begin{eqnarray}\label{coeff}
c&=&-4\frac{G}{G_N}\nonumber\\
d&=&\frac{G}{G_N}\frac{2-a_1-\lambda(4-3a_1)}{2(1-\lambda)}\nonumber\\
\gamma&=&\frac{G}{G_N} a_1-\frac{a_2}{a_1},
\end{eqnarray}
where $G_N$ is the Newton constant, that could in principle differ from $G$, introduced through the parameter $\zeta$ in the Horava-Lifshitz action, see eq.(\ref{HLaction}).\\
Describing the source in terms of the energy density $\rho$ and velocity ${\bf v}$, the potentials appearing in the solution are defined as follows:
\begin{eqnarray*}
U&\equiv&\int\frac{\rho({\bf x'},t)}{|{\bf x}-{\bf x'}|}d^3x',\\
\chi&\equiv&\int\rho({\bf x'},t)\ |{\bf x}-{\bf x'}|\ d^3x'\\
V_j&\equiv&\int\frac{\rho({\bf x'},t)\ v'_j}{|{\bf x}-{\bf x'}|}\ d^3x'.
\end{eqnarray*}

From field equations, one gets
\begin{eqnarray*}
\partial^2 U&=&-4\pi G_N\rho,\\
\partial^2V_j&=&-4\pi G_N\rho v_j
\end{eqnarray*}
 
 Moreover, from the continuity equation for the source, one obtains $ \chi_{,0i}=V_i-W_i$, where the potential $W_i$ is now defined
 as
 $$
 W_i\equiv\int\frac{\rho({\bf x'},t)\ {\bf v' \cdot  (x-x')}(x-x')_i}{|{\bf x}-{\bf x'}|^3}d^3x'.\\
 $$
 
 Then, variation from General Relativity solutions arise because the perturbed solution depends on the parameters $c$, $d$ and $\gamma$ that in turn  explicitly depend on the Horava-Lifshitz couplings.
 
\begin{figure}
 \includegraphics[width=9cm]{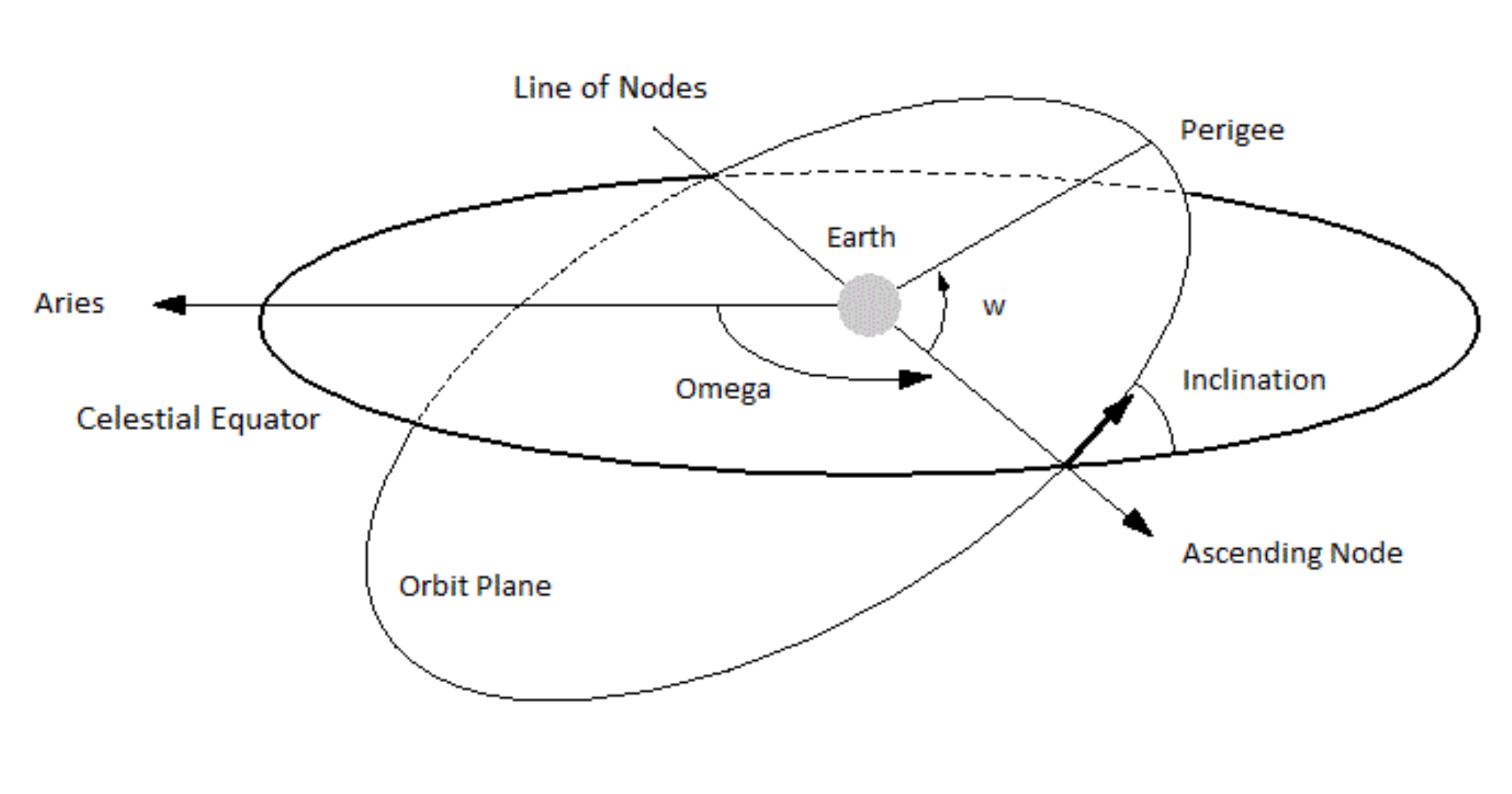}
 \caption{Keplerian orbital parameters. The longitude of the ascending node  $\Omega$  is defined to be the angle between a stationary reference line, the vernal equinox which is also called the first point of Aries, and the line connecting the origin of the coordinate system and the point where the orbiting body intersects the orbital plane as it is moving upwards \cite{brouwer61}.}\label{orbitalelem}
 \end{figure}

 With this solution in mind, let us specify the source terms for the Earth, considered as a massive homogeneous spherical body with mass $m$ and intrinsic spin-angular momentum ${\bf  J}=(2/5)mR^2 \boldsymbol{\omega}$, being $R$ its radius and ${\boldsymbol \omega}$ its angular velocity. It is considered at rest in the origin of the axes. In this approximation the vector potential $V_i$ and $W_i$ coincide and read \cite{will}
 $$
 V^i=W^i=\frac{1}{2}\left({\bf n}\times\frac{ {\bf J}}{r^2}\right)^i,
 $$
  where $r$ is the distance from the Earth and $n^i=x^i/r$ are the components of the  unit vector. In such a case the dependence of the solution on the theory parameters reduces to the ratio $G/G_N$ and $\gamma$ as it can be seen from eqs. (\ref{PPN}) and (\ref{coeff}) and the fact that $\chi_{,0i}=0$.\\
 Let us evaluate such a contribution. We will use the gravito-magnetic formalism, thus introducing the vector potential 
 $$
 A_{\mu}=-\frac{1}{4}\bar{h}_{0\mu},
 $$
with
$$
\bar{h}_{\mu\nu}=h_{\mu\nu}-\eta_{\mu\nu} h/2
$$
 being the trace-reversed metric perturbation. Then the gravitomagnetic field can be introduced ${\bf B}={\bf \nabla\times A}$ and the contribution coming from a spherical rotating homogeneous sphere can be easily calculated. It turns out to be equivalent to the General Relativity field except for the factor $G/G_N$. Choosing ${\bf J}$ to lie perpendicular to the celestial equator (see fig.\ref{orbitalelem}), one infers
 \begin{equation}\label{gravitomag}
 {\bf B}_{GR}=-\frac{1}{2}\frac{{\bf J}}{r^3}+\frac{3}{2}\frac{({\bf J\times \hat{r}})\ {\bf \hat{r}}}{r^3}
 \end{equation}
 and 
  \begin{equation}
 {\bf B}_{HL}= \frac{G}{G_N} \  {\bf B}_{GR}.
 \end{equation}
 In order to investigate how this field will affect the motion of test particles around the Earth, we will use the Gaussian perturbation equations \cite{brouwer61} that give the time variation of the Keplerian orbital elements in the presence of a perturbing force. In the case at hand, we treat the gravitomagnetic force ${\bf F}=-4{\bf v \times B}$ as a small perturbation of the otherwise unperturbed Keplerian motion. We here focus only on the secular variation of the longitude of the ascending node $\Omega$ that is defined to be the angle between a stationary reference line and the line connecting the origin of the coordinate system and the point where the orbiting body intersects the orbital plane as it is moving upwards, see fig.\ref{orbitalelem}.\\
The time variation of $\Omega$ is connected in General Relativity with the Lense-Thirring drag \cite{lense18}. In particular, averaging over one period one can separate the secular variation of the perturbation that in General Relativity reads
\begin{equation}
\langle\dot{\Omega}_{GR}\rangle=\frac{2  G_N J}{\text{a}^3(1-e^2)^{3/2}},
\end{equation}
being '$\text{a}$' the semimajor axis of the orbit of the test body and $e$ the orbit eccentricity. The ratio between the Horava-Lifshitz and the General Relativity contributions does not depend on the details of the orbital configuration since their functional dependence is the same. It only constraints the coupling constant $G$ that appears in eq.(\ref{HLaction}). Actually, the measurement of $\dot{\Omega}$ by space orbiting experiments and its confrontation with the General Relativity estimation
\begin{equation}\label{lense}
\frac{\langle\dot{\Omega}_{HL}\rangle}{\langle\dot{\Omega}_{GR}\rangle}=\frac{G}{G_N}
\end{equation}
gives an upper bound on the ratio on the coupling constants $G$ and $G_N$. 

On the other hand,  the gravitomagnetic and the gravitoelectric fields will also cause a precession of the spin $\bf{S}$ of gyroscopes orbiting around the Earth. A gyroscope will undergo precession due to two torques. One is known as the geoedetic precession \cite{desitter16} and is independent of the Earth gravitomagnetic field; the other is due to a coupling to the gravitomagnetic field:
\begin{equation}
\dot{\bf{S}}={\bf{\Omega}}\times{\bf{S}},\quad\quad{\bf\Omega}={\bf\Omega^G}+{\bf\Omega^{LT}},
\end{equation}
where ${\bf\Omega^G}$ and ${\bf\Omega^{LT}}$ are the geodetic and Lense-Thirring precession, respectively.\\
The value of this variation in Horava-Lifshitz theory can be deduced in terms of the weak field solution written above; in particular, the Lense-Thirring term has been evaluated above and is related to the General Relativity value by means of eq.(\ref{lense}). For what concerns the de Sitter (geodetic) effect, the General Relativity contribution is
$$
{\bf\Omega}^{G}_{GR}=\frac{3}{2 r^3}GM({\bf r} \times {\bf v})
$$
and ratio between the Horava-Lifshitz value and the General Relativity one turns out to be
\begin{equation}\label{geodetic}
\frac{\Omega^{G}_{HL}}{\Omega^{G}_{GR}}=\frac{1}{3}\left(1+2\frac{G}{G_N}a_1-2\frac{a_2}{a_1}\right)
\end{equation}
We note that the geodesic effect allows us to constrain the matter couplings $a_1$ and $a_2$ even though only trough the combination shown above. \\ Because of the peculiarity of the situation analysed here (in our case $V_i=W_i$ as mentioned before eq.(\ref{gravitomag})),  both effects do not depend on the parameter $\lambda$ that is the coefficient of the extrinsic curvature term.

%%%%%%%%%%%%%%%%%%%%%%%%%%%%%%%%%%%%%%%%%%%%%%%%%%%%%%%%%%%%%%%%%%%%%%%%%%%%%%%%%%%%%%%%%%%%%%%%%%%%%%%%%%%%%%%%%%%%%%%%%%%%%%%%%%%%%%%%%%%%%%%%%%%%%%%%%%%%%%%%%%%%%%%%%%%%%%%%%%%%%%%%%%%%%%%%%%%%%%%%%%%%%%%
\section{Constraints  from space experiments}\label{Constraints  from space experiments}
The Gravity Probe B (GP-B) four gyroscopes aboard an Earth-orbiting satellite \cite{gpb} allowed to measure the frame-dragging effect with an error of about $19\%$  \cite{everitt11} $\Omega ^{LT}_{obs}=37.2\pm 7.2 \ \ \text{mas/yr}$,  while the General Relativity predicted value is $\Omega^{LT}_{GR}=39.2 \ \ \text{mas/yr}$. This result provides a constraint on the difference between the coupling constant $G$ that appears in the Horava-Lifshitz theory and the Newton constant $G_N$.
Indeed, 
$$
\left|\frac{\Omega^{LT}_{obs}-\Omega^{LT}_{GR}}{\Omega^{LT}_{GR}}\right|=0.05\ \ \ \Longrightarrow\ \ \ \left|\frac{G}{G_N}-1\right|=0.05.
$$
The mission was able to measure the geodetic effect as well. The measured value reads $\Omega^{G}_{obs}=-6601.8\pm18.3\ \ \text{mas/yr}$, at the level of $0.28\%$. The General Relativity predicted value turns out to be $\Omega^{G}_{GR}=-6606.1\ \ \text{mas/yr}$ . In this case the observations constraint a combination of parameters, namely
$$
\left|\frac{\Omega^{G}_{obs}-\Omega^{G}_{GR}}{\Omega^{G}_{GR}}\right|=\left|\frac{2}{3}\left(\frac{G}{G_N}a_1-\frac{a_2}{a_1}-1\right)\right|=0.0006
$$

The LAGEOS  satellites (LAser GEOdynamics Satellites), launched by NASA (LAGEOS) and NASA-ASI (LAGEOS-2) \cite{lageos}, have been recently able to test the Lense-Thirring effect. This mission has  improved the sensitivity  thanks to the laser-ranging technique for measuring distances as well as the combination of the two LAGEOS nodal longitudes that allow to eliminate the uncertainty in the value of the second degree zonal harmonic describing the Earth's quadrupole moment. \\
The Lense-Thirring effect measured for the combination of the two LAGEOS nodal longitudes reads \cite{ciufolini04}
$$
\dot\Omega^{LT}_{obs}=47.9 \ \ \text{mas/yr},
$$ 
with an error of about $10\%$. In this case the value predicted by General Relativity is $\dot\Omega^{LT}_{GR}=48.2 \ \ \text{mas/yr}$; then, the constraint on the parameter c reduces to
$$
\left|\frac{G}{G_N}-1\right|=0.006.
$$
Lastly, the Laser Relativity Satellite (LARES) mission \cite{lares}, launched on february 2012 and founded by ASI, aims to achieve an uncertainty of a few percent only. Actually,  the three nodes of the LAGEOS, LAGEOS-2 and LARES satellites together with gravitational field determinations from the GRACE space mission (Gravity Recovery And Climate Experiment)  \cite{grace} will allow to improve the previous results by eliminating the uncertainties in the value of the first two even zonal harmonics of the Earth potential.\\
 At the present stage, Monte Carlo simulations predict a standard deviation of the  simulated value of the frame-dragging to be equal to $1.4\%$ of the frame-dragging effect predicted by General Relativity and its mean value effect is equal to $100.24\%$ if its general relativistic value \cite{ciufolini13}.\\
 This prediction would constraint the combination of the Horava-Lifshitz coupling constant $G$ and the Newton constant $G_N$ to differ from unity by $2\cdot10^{-3}$:
$$
\left|\frac{G}{G_N}-1\right|=0.002.
$$
It could be instructive to compare these constraints with the results coming from experimental tests.\\ Among the fundamental constants, the gravitational constant $G_N$ is the  less accurately measured. Newton's law of gravitation has been used to test it at laboratory scales but also at geophysical and astronomical scales. Nevertheless, the latter mainly gives information only on $GM$: through the measurements of planetary and satellite orbits only the product of the Newton's constant times the masses of the interacting bodies can be directly constrained. \\
Quite recently in \cite{umezu05} an independent estimation has been derived. In that work the spatial variation of the gravitational constant has been parametrized by $G(r)=\xi G_N$\footnote{$\xi$ is related to the parameter $c$ of the Horava Lifshitz theory: $\xi=c/4$}. At different scales, a change of the Newton constant affects various phenomena. For scales up to $r\sim 10^{10}m$ the analysis of the age of  stars in globular clusters constraints $\xi$, since an increasing Newton's constant causes stars to burn faster. The value obtained in \cite{umezu05} is  $0.93\lesssim\xi\lesssim 1.09$.\\
Further, the primordial light-element abundances constraint the value of the Newton's constant during the BBN epoch. In particular, the Helium abundance and the Deuterium-to-hydrogen abundance give $0.95\lesssim\xi\lesssim1.01$ at $10^{8}\div10^{12} m$ scales.\\
Finally,  on cosmological scales, a changing Newton's constant appears in the amplitude of the acoustic peaks in the CMB power spectrum, even though it provides a weaker constraint. By using WMAP data the range, at $95\%$ confidence level, is $0.75\lesssim\xi\lesssim1.66$.

%%%%%%%%%%%%%%%%%%%%%%%%%%%%%%%%%%%%%%%%%%%%%%%%%%%%%%%%%%%%%%%%%%%%%%%%%%%%%%%%%%%%%%%%%%%%%%%%%%%%%%%%%%%%%%%%%%%%%%%%%%%%%%%%%%%%%%%%%%%%%%%%%%%%%%%%%%%%%%%%%%%%%%%%%%%%%%%%%%%%%%%%%%%%%%%%%%%%%%%%%%%%%%%
\section{Terrestrial experiment: GINGER}
GINGER (Gyroscopes IN GEneral Relativity) is a proposed tridimensional array of laser gyroscopes with the aim of measuring general relativity effects in a terrestrial laboratory by means of laser gyroscopes \cite{ginger}. Being the whole proposed experiment Earth based, it has a constant gravito-electric field so it does not require any modelling of the interior of the Earth. 

The experiments in space are based on the precession of physical gyroscopes induced by the gravitational field so that the coupling of the angular momentum of a gyroscope with the gravitational field induces a torque depending on the configuration of the field. \\GINGER experiment, instead, exploits the anisotropic propagation of light in the skew symmetric space-time associated with rotating bodies \cite{tartaglia14}. \\
The key idea of the GINGER experiment is to measure the difference in the times of flight of two beams circulating in a laser cavity in opposite directions. This translates into a time difference $\delta\tau=\tau_{+}-\tau_{-}$ between the right-handed beam propagation time ($\tau_{+}$) and the left-handed one ($\tau_{-}$)
\begin{equation}
\delta\tau=-2\sqrt{g_{00}}\oint\frac{g_{0i}}{g_{00}}ds^i.
\end{equation}
The presence of a gravitoelectric (Newtonian) field  as well as a gravitomagnetic contribution due to the rotation of the Earth. Following the analysis in \cite{bosi11}, the General Relativity calculation, in linear approximation for an instrument with its normal contained in the local meridian plane, gives
\begin{eqnarray}
c\ \delta\tau&=&\frac{4A}{c}\Omega_{E}\left[\cos{(\theta+\alpha)}-\frac{G_NM}{c^2 R}\sin{\theta}\sin{\alpha}\right.\nonumber\\
&-&\left.\frac{G_NI_E}{c^2 R^3}(2\cos{\theta}\cos{\alpha}+\sin{\theta}\sin{\alpha})\right],
\end{eqnarray}
where $A$ is the area encircled by the light beams, $\alpha$ is the angle between the local radial direction and the normal to the plane of the array-laser ring, measured in the meridian plane, and $\theta$ is the colatitude of the laboratory; $\Omega_{E}$  ($I_E$) is the rotation rate  (momentum of inertia) of the Earth as measured in the local reference frame. The first constribution in the square brackets is the Sagnac term, due to the rotation of the Earth; the other terms encode the relativistic effects: geodetic and Lense-Thirring term\footnote{Actually, also the Thomas precession, related to the angular defect due to the Lorentz boost comes into play, but it is two orders of magnitude smaller than the geodetic and Lense-Thirring terms.}. To cancel out the pure kinematic term, an accuracy $1$ in $10^{10}$ is necessary. This can be achieved, according to the proposal, with a detector consisting of six large ring lasers arranged in three orthogonal axes and in about two years of measurements.

For a theory that differs from General Relativity but is still in the set of metric theories that can be described by the PPN formalism, the contribution to the time shift has been evaluated in \cite{bosi11} and, for the Horava-Lifshitz theory, the geodesic and Lense-Thirring contributions read:

\begin{eqnarray}
c\ \delta\tau&=&\frac{4A}{c}\Omega_{E}\left[-\left(1+\frac{G}{G_N} a_1-\frac{a_2}{a_1}\right)\frac{G M}{c^2 R}\sin{\theta}\sin{\alpha}\right.\nonumber\\
&-&\left.\frac{G I_E}{c^2 R^3}\left(2\cos{\theta}\cos{\alpha}+\sin{\theta}\sin{\alpha}\right)\right].
\end{eqnarray}
As for the GP-B mission, a measurement of the geodetic and Lense-Thirring effect would give constraints on the parameters of the Horava-Lifshitz theory.
%%%%%%%%%%%%%%%%%%%%%%%%%%%%%%%%%%%%%%%%%%%%%%%%%%%%%%%%%%%%%%%%%%%%%%%%%%%%%%%%%%%%%%%%%%%%%%%%%%%%%%%%%%%%%%%%%%%%%%%%%%%%%%%%%%%%%%%%%%%%%%%%%%%%%%%%%%%%%%%%%%%%%%%%%%%%%%%%%%%%%%%%%%%%%%%%%%%%%%%%%%%%%%%
\section{Conclusions}\label{Conclusions}
In this paper we have considered the weak field and slow motion limit of a covariant version of Horava-Lifshitz theory in order to constrain the parameters of the model against the results of some recent space experiments. Such an approximation is well suited for describing the field around the Earth (massive and slowly rotating body). 

Satellites orbiting around our planet experience a weak field induced dynamics and move according to the slow motion approximation. Parametrized Post Newtonian formalism is a good framework to deal with in such a situation. 

Going into details, when considering the solar system scenario, all the terms that come from the cosmological constant and the space curvature are negligible and the solution explicitly depends only on a few parameters. In particular, at the third post-newtonian order, the potentials depend on $G$, the Newtonian constant in the Horava-Lifshitz theory, on $\lambda$, the coefficient of the extrinsic curvature term  that characterises deviations of the kinetic part of the action from General Relativity, as well as on $a_1$ and $a_2$ that are two arbitrary coupling constants in the matter action.\\
The difference between General Relativity and Horava-Lifshitz solution translates into different prediction on the motion of satellites around the Earth as well as on the precession of gyroscopes. 

In particular we have deduced the Lense-Thirring precession and the de Sitter effect. The former only depends on the ratio $\frac{G}{G_N}$, that is the difference on the coupling constants $G$ and $G_N$. The latter introduces a constraint on the matter couplings $a_1$ and $a_2$ through the combination $\frac{1}{3}\left(1+2\frac{G}{G_N}a_1-2\frac{a_2}{a_1}\right)$.\\
We then have also considered constraints from observational data. The Gravity Probe B experiment provides a measurement of the frame dragging  with an error of about $19\%$ and the geodetic effect with an error of $0.28\%$. The two induce constraints on $G$, that is forced to be equal to the Newton constant $G_N$ up to five parts over $10^2$ and on the combination $\left(\frac{G}{G_N}a_1-\frac{a_2}{a_1}\right)$ that must be equal to unity.\\ 
When considering Lense-Thirring effect from Lageos experiment, which reached an accuracy of $10\%$, the constraint on $G$ increases of about an order of magnitude and the Monte Carlo simulations on LARES seem to predict the ratio $G/G_N$ to differ from unity by $2\cdot 10^{-3}$. \\
Finally, we have presented the theoretical result expected by GINGER, an Earth based experiment that aims to evaluate the response to the gravitational field of a ring laser array. Again, from GINGER forthcoming data we will get a combination of the de Sitter and Lense-Thirring effect thus providing independent constraints on the same parameters of the Horava-Lifshitz theory.

%%%%%%%%%%%%%%%%%%%%%%%%%%%%%%%%%%%%%%%%%%%%%%%%%%%%%%%%%%%%%%%%%%%%%%%%%%%%%%%%%%%%%%%%%%%%%%%%%%%%%%%%%%%%%%%%%%%%%%%%%%%%%%%%%%%%%%%%%%%%%%%%%%%%%%%%%%%%%%%%%%%%%%%%%%%%%%%%%%%%%%%%%%%%%%%%%%%%%%%%%%%%%%%
\section*{Acknowledgements}
Authors wish to thank Agenzia Spaziale Italiana (ASI)  for partial support through contract n.I/034/12/0  and Istituto nazionale di Fisica Nucleare (INFN).

%%%%%%%%%%%%%%%%%%%%%%%%%%%%%%%%%%%%%%%%%%%%%%%%%%%%%%%%%%%%%%%%%%%%%%%%%%%%%%%%%%%%%%%%%%%%%%%%%%%%%%%%%%%%%%%%%%%%%%%%%%%%%%%%%%%%%%%%%%%%%%%%%%%%%%%%%%%%%%%%%%%%%%%%%%%%%%%%%%%%%%%%%%%%%%%%%%%%%%%%%%%%%%%

%%%%%%%%%%%%%%%%%%%%%%%%%%%%%%%%%%%%%%%%%%%%%%%%%%%%%%%%%%%%%%%%%%%%%%%%%%%%%%%%%%%%%%%%%%%%%%%%%%%%%%%%%%%%%%%%%%%%%%%%%%%%%%%%%%%%%%%%%%%%%%%%%%%%%%%%%%%%%%%%%%%%%%%%%%%%%%%%%%%%%%%%%%%%%%%%%%%%%%%%%%%%%%%

\begin{thebibliography}{99}

\bibitem{will} C.M. Will, Living Rev. Relativity 17,  (2014),  4. URL (cited on 1/07/2014): http://www.livingreviews.org/lrr-2014-4.

\bibitem{ruggiero07}  M.L. Ruggiero and L.Iorio, JCAP {\bf 07} (2007) 010.

\bibitem{smith08} T.L. Smith, A.L. Erickcek, R.R. Caldwell and M. Kamionkowski, Phys. Rev. D {\bf 77}, 024015 (2008).

\bibitem{bohmer10} C.G. B\"ohmer, G. De Risi, T. Harko and F. S.N. Lobo, Class. Quantum Grav. {\bf 27} (2010) 185013.

\bibitem{harko11} T. Harko, Z. Kov\'acs and F.S.N. Lobo, Proc. R. Soc. A (2011) {\bf 467}, 1390-1407.

\bibitem{lambiase13} G. Lambiase, M. Sakellariadu and A. Stabile, 	JCAP {\bf 12}(2013) 020.

\bibitem{Horava09} P. Horava, Phys. Rev. D {\bf 79}, 084008 (2009).

\bibitem{dirac} Dirac, P.A.M. Proceedings of the Royal Society of London, Mathematical and Physical Sciences {\bf 46}, Issue 1246, pp 333-343 (August 1958).

\bibitem{arnowitt59} Arnowitt, R., Deser, S., Misner, C., Phys. Rev. {\bf 116} (5): 1322Ð1330 (1959).

\bibitem{wang11} Anzhong Wang and Yumei Wu, Phys. Rev. D {\bf 83}, 044031 (2011).

\bibitem{HL} P.Horava and C. Melby-Thompson, Phys. Rev. D {\bf 82}, 064027 (2010).

\bibitem{daSilva11} A. M da Silva, Class.Quant.Grav. {\bf 28}, 055011 (2011).

\bibitem{willbook} C.F. Will, {\sl Theory and experiments in gravitational physics}, Cambridge: Cambridge University Press, 1981.

\bibitem{desitter16} W. de Sitter, Mon. Not. Roy. Astron. Soc. {\bf 77}, 155--184 (1916).

\bibitem{lense18} J. Lense and H.Thirring,  Phys. Z. {\bf 19}, 156--163 (1918).

\bibitem{wheeler} I. Ciufolini and J.A. Wheeler, {\sl Gravitation and Inertia}, Pronceton University Press, 1996.

\bibitem{gpb} {\sl http://einstein.stanford.edu/}.

\bibitem{lageos} {\sl http://science.nasa.gov/missions/lageos-1-2/}.

\bibitem{lares} {\sl http://www.lares-mission.com/}.

\bibitem{ginger} http://www.df.unipi.it/ginger

\bibitem{tartaglia14}  A. Tartaglia {\sl et al.}, EPJ Web Conf. {\bf 74} 03001, 2014.

\bibitem{lin13} Kai Lin, Shinji Mukohyama, Anzhong Wang and Tao Zhu, Phys. Rev. D {\bf89}, 084022 (2014).

\bibitem{SVW09} T. Sotiriou, M. Visser and S. Weinfurtner, JHEP {\bf 10}, 033 (2009).

\bibitem{lin12} Kai Lin, Shinji Mukohyama and Anzhong Wang, Phys. Rev. D {\bf 86}, 104024 (2012).

%\bibitem{huang12} Yongqing Huang and Anzhong Wang, JCAP {\bf 10} 010 (2012).

%\bibitem{wang10} Anzhong Wang, Phys. Rev. D {\bf 82}, 124063 (2010).

%\bibitem{greenwald10} J. Greenwald, V.H. Satheeshkumar and A. Wang, JCAP {\bf 12} 007 (2010).

%\bibitem{alexandre11} J. Alexandre, P. Pasipoularides, Phys. Rev. D {\bf 83} 084030 (2011).



%\bibitem{wang12}Anzhong Wang, Phys. Rev. Lett. {\bf 110}, 091101 (2013).

\bibitem{brouwer61} D. Brouwer and G.M. Clemece, {\sl Methods of Celestial Mechanics}, Academic Press, New York, 1961.

\bibitem{everitt11} C.W. Everitt {\sl et al.}, Phys. Rev. Lett. {\bf 106}, 221101 (2011).

\bibitem{ciufolini04} I. Ciufolini and E.C. Pavlis, Nature {\bf 431}, 958--960 (2004).

\bibitem{grace} {\sl http://www.csr.utexas.edu/grace/}.

\bibitem{ciufolini13} I. Ciufolini {\sl et al.}, Class.Quant.Grav. {\bf 30} (2013) 235009
 .

\bibitem{umezu05} Ken-ichi Umezu, Kiyotomo Ichiki and Masanobu Yahiro, Phys. Rev. D {\bf 72}, 044010 (2005).

\bibitem{bosi11} F. Bosi {\sl et al.}, Phys. Rev. D {\bf 84}, 122002 (2011).



\end{thebibliography}
\end{document}